\documentclass[conference]{IEEEtran}

\usepackage{cite}

\usepackage{amsmath,amssymb,amsfonts}
\usepackage{algorithmic}
\usepackage{graphicx}
\usepackage{textcomp}
\usepackage{xcolor}
\usepackage{hyperref}
\usepackage{siunitx}
\usepackage{svg}
\usepackage{pgfplots}

\usepackage{glossaries}
\makeglossaries
\newacronym{mimo}{MIMO}{Multiple Input Multiple Output}
\newacronym{pa}{PA}{Power Amplifier}
\newacronym{bs}{BS}{Base Station}
\newacronym{ue}{UE}{User Equipment}
\newacronym{ib}{IB}{in-band}
\newacronym{oob}{OOB}{Out-of-Band}
\newacronym{rts}{RTS}{Ray-Tracing Simulator}
\newacronym{los}{LoS}{Line of Sight}
\newacronym{nlos}{NLoS}{Non Line of Sight}
\newacronym{evm}{EVM}{Error Vector Magnitude}
\newacronym{kpi}{KPI}{Key Performance Indicator}
\newacronym{qam}{QAM}{Quadrature Amplitude Modulation}
\newacronym{ofdm}{OFDM}{Orthogonal Frequency-Division Multiplexing}
\newacronym{mr}{MR}{Maximum Ratio}
\newacronym{zf}{ZF}{Zero Forcing}
\newacronym{aclr}{ACLR}{Adjacent Channel Leakage Power Ratio}
\newacronym{bo}{BO}{back-off}
\newacronym{psd}{PSD}{Power Spectral Density}
\newacronym{im}{IM}{Intermodulation}
\newacronym{arp}{ARP}{Antenna Reference Point}
\newacronym{ag}{AG}{Array Gain}
\newacronym{ula}{ULA}{Uniform Linear Array}
\newacronym{rrc}{RRC}{Root Raised Cosine}
\newacronym{ecdf}{ECDF}{Empirical Cumulative Distribution Function}

\def\BibTeX{{\rm B\kern-.05em{\sc i\kern-.025em b}\kern-.08em
    T\kern-.1667em\lower.7ex\hbox{E}\kern-.125emX}}

\newcommand\blfootnote[1]{%
  \begingroup
  \renewcommand\thefootnote{}\footnote{#1}%
  \addtocounter{footnote}{-1}%
  \endgroup
}

\begin{document}
\title{Out-of-Band Distortion in Massive MIMO: \\
    What to expect under realistic conditions?
}

\author{\IEEEauthorblockN{Laura Monteyne\IEEEauthorrefmark{1},
		Gilles Callebaut\IEEEauthorrefmark{1},
        Björn Sihlbom\IEEEauthorrefmark{2} and		
        Liesbet Van der Perre\IEEEauthorrefmark{1}
		}

\IEEEauthorblockA{\IEEEauthorrefmark{1}Department of Electrical Engineering, KU Leuven, Belgium}

\IEEEauthorblockA{\IEEEauthorrefmark{2}Research Center at Huawei Technologies, Gothenburg, Sweden}

\IEEEauthorblockA{E-mail: laura.monteyne@kuleuven.be}

}

\maketitle

\begin{abstract}
Massive \gls{mimo} offers superior capacity for future networks. In the quest for energy efficient implementation of these large array-based transmission systems, the power consumption of the \glspl{pa} is a main bottleneck. This paper investigates whether it is possible to operate the \glspl{pa} in their efficient nonlinear region, as the \gls{oob} distortion may not get the same array gain as the \gls{ib} signals. We present a framework to simulate the effects under realistic conditions, leveraging on an accurate \gls{rts}. The results show that the often assumed i.i.d. Rayleigh fading channel model results in too optimistic predictions, also in \gls{nlos} multi-path scenarios, regarding the spatial distribution of \gls{oob} emissions. We further comment on the consequences in view of current regulatory constraints. 
\end{abstract}


\section{Introduction}
Massive \gls{mimo} is a key technology to achieve high spectral efficiency and capacity gain for the 5G network, especially in dense urban environments. The challenge is to build such large systems to process a \textit{massive} number of antenna signals at reasonable complexity. The need for energy efficiency, low complexity and thus low-cost hardware components is evident. In modern cellular networks, the \gls{bs} and more specifically its \glspl{pa}, are the biggest energy consumers~\cite{danve}. Several studies suggest the possibility of implementing massive \gls{mimo} systems with low complexity hardware, while maintaining the high performance~\cite{bjornson2014, Desset}. Unfortunately, a reduced complexity in the \gls{pa} hardware results in higher nonlinearities and thus in \gls{ib} and \gls{oob} distortions~\cite{yadav}. \gls{oob} distortions are defined by ITU-R \cite{ITU} as emissions on a frequency immediately outside the necessary frequency bandwidth and could affect neighbouring applications. The analysis in this paper focuses on \gls{oob} distortions, while it is considered that \gls{ib} performance should be guaranteed at all times.

\blfootnote{This paper is accepted and published in the Conference Proceedings of 2021 IEEE 94th Vehicular Technology Conference (2021).}

Several studies on the impact of nonlinear amplification in massive \gls{mimo} transmission have been conducted. They are mostly based on relatively simple models for both channels and distortions, such that it is possible to derive closed form analytical expressions. Authors have assumed that distortions due to nonlinear amplification can be considered uncorrelated in i.i.d. Rayleigh fading conditions. Consequently, their impact highly reduces with increasing number of antennas~\cite{bjornson2014}. However, in many cases the uncorrelated terms assumption is not valid and for example, it has been shown that in \gls{los} situations with a single or dominant user, the \gls{oob} distortion will get an array gain in the direction of the intended user~\cite{Larsvdp2018}. The directivity of \gls{oob} distortions due to nonlinear amplification has been further studied for multi-user scenarios~\cite{Mollen2018}, where also \gls{im} beams~\cite{arraybible} are predicted. In the published state-of-the-art, omnidirectional antennas have been typically assumed, which are practically not feasible nor desirable in view of mutual coupling.
In summary, the state-of-the-art  presents different theoretical results that leave the question open what to expect under realistic conditions. To draw relevant and reliable conclusions, realistic models and assumptions are required. 

The main objective of this study is to investigate and analyze the impact of low-power transmitters, based on low-cost hardware, on the \gls{oob} distortion in a massive \gls{mimo} system under realistic conditions. Do nonlinear \glspl{pa} in massive \gls{mimo} cause correlated \gls{ib} and \gls{oob} distortions and under what circumstances are these distortions problematic with respect to current regulations? We specifically investigate the spatial distribution of \gls{oob} distortions in massive \gls{mimo}.

The results in this paper are generated by a versatile yet robust simulation framework for the analysis of \gls{oob} radiation in a massive \gls{mimo} environment under realistic conditions. In particular, channel modeling in the simulation framework is realized with advanced \gls{rts} software, provided by Huawei Technologies Sweden. 
Several evaluation metrics are defined and graphical heatmaps for the spatial distribution of the \gls{oob} radiation are presented to quantify and measure the impact of the nonlinear characteristics of the \glspl{pa}. We study both random scenarios and `handpicked' use cases that are created to provoke potentially worst case impact in terms of \gls{oob} emission. Finally, we analyze and interpret the results. 

The second section introduces the simulation framework that is established with the purpose of modeling and analyzing \gls{oob} emission in massive \gls{mimo} transmission. We define the evaluation metrics that are used to compare and assess different scenarios.  In Section~\ref{sec:analysis}, we describe the set-up for the performed experiments to analyze \gls{oob} distortion. 
The results of this analysis are presented and discussed in Section~\ref{sec:results}. Section~\ref{sec:regulations} discusses the current \gls{oob} regulations in view of the spatial distribution of \gls{oob} occurring from massive \gls{mimo} transmission. Finally, the main conclusions are summarized in section \ref{sec:conclusion}, along with future work.

\section{Simulation framework}
\begin{figure}[!htb]
    \centering
    \includegraphics[width=\linewidth]{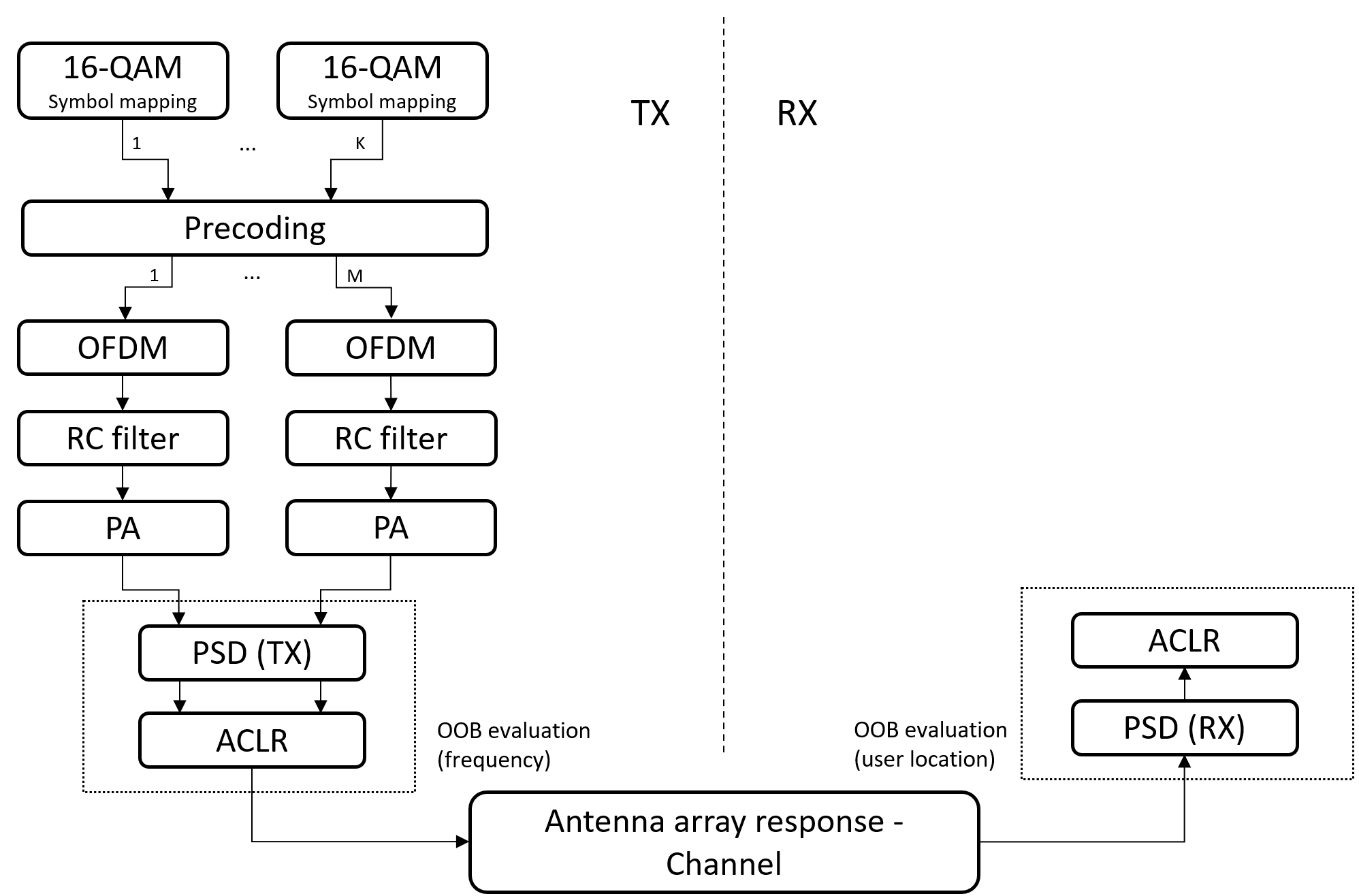}
    \caption{Overview of the simulation framework.}
    \label{fig:framework}
\end{figure}

A high-level overview of the simulation framework is given in Fig.~\ref{fig:framework}.
The first building blocks generate random data and maps it to the 16-\gls{qam} constellation. The symbols are modulated to an \gls{ofdm} signal based on the LTE-format. Pulse shaping is realized with a root raised cosine filter with a roll-off factor $\alpha_{\text{RRC}} = 0.22$. \Gls{mr} and (regularized) \gls{zf} precoders are implemented. The symbols are oversampled with a factor 4 in the time domain. Following amplification, possibly inducing nonlinear distortion, the signals are transmitted via the channels and further considered both at the locations of the intended users and observer points in the coverage area. 
In the following, we discuss the channels generated by means of the \gls{rts} software and the implemented nonlinear \gls{pa} model. Furthermore, we show what evaluation metrics are available in the framework to analyze a given scenario.

\subsection{Channel generation and PA modeling}
Channel models are realized based on \gls{rts} software, implementing highly realistic and practical models based on the  channel models developed in the METIS project~\cite{Metis}. A particular strength of the simulator is that it is able to trace a large number of rays. It is able to capture absorption and reflection phenomena, and generates detailed channel responses including reflections, diffraction and diffuse scattering phenomena.
Any area, of particular interest being dense urban surroundings, can be selected via Open Street Map\footnote{\url{www.openstreetmap.org} is an open license map of the world. The selected area can be added by means of its coordinates.} and added to the framework. The height of buildings can be altered in the extracted data, if necessary. Antennas and arrays of selected size can be implemented by feeding their pattern to the simulator. The number of antennas at the \gls{bs} is hereby denoted as $M$. 
The antennas can be placed on a location of preference. Users can be assigned a specific antenna and location as well. The number of \glspl{ue} is denoted as $K$. One of the outputs of this simulator are the realistic massive \gls{mimo} channel responses with dimension $[M\times K\times f]$ with $f$ for frequency. Note that \gls{oob} frequencies are considered as well. We are investigating the \SI{3.5}{\giga\hertz}~band, which is currently worldwide the most appropriate for TDD-based massive MIMO, with a bandwidth $B$ equal to \SI{40}{\mega\hertz}. 
The nonlinear behavior of the \gls{pa} is modeled with a third-order nonlinear term in function of the input signal $s$:
\begin{equation}
    s_{pa} = s_{norm} - \alpha \cdot s_{norm}^3,
\end{equation}
where $\alpha = 1-10^{-1/20}$. The normalization factor $s_{norm}$ is calculated as:
\begin{equation}
    s_{norm} = \frac{s}{\sigma_s \cdot 10^{\text{BO}/20}},
\end{equation}
with $\sigma_s$ the standard deviation of input signal $s$, in other words the power of the signal. The input power \gls{bo} of the amplifier, expressed in dB, is an input parameter of this \gls{pa} model. So, in our model we can specify how far from the saturation point we want the \gls{pa} to operate and thus manipulate the level of nonlinearity caused by the \gls{pa}.

\subsection{Performance evaluation metrics}
The following \glspl{kpi} are defined and implemented in our simulation framework, to be able to objectively analyze different scenarios. We calculate the \gls{psd} in Python with Welch's method~\cite{welch}. It is expressed in dB/Hz and calculated for \gls{ib} as well as \gls{oob} frequencies. Based on~\cite{mollen}, we calculate the \gls{aclr} in a \gls{mimo} environment as follows:
\begin{equation}
    ACLR \triangleq \frac{\max \left\{\int_{-3B/2}^{-B/2} \mathrm{PSD}(f) \mathrm{d} f, \int_{B / 2}^{3 B / 2} \mathrm{PSD}(f) \mathrm{d} f\right\}}{\int_{-B/2}^{B/2} \mathrm{PSD}(f) \mathrm{d} f},
\end{equation}
which is expressed in dB. The numerator selects the maximum of respectively the left and right adjacent frequency band, both with bandwidth $B$. A graphical result is available in the \gls{oob} heatmap where the \gls{oob} power in dB is indicated by a colored scale on top of the geographical area (as in the example given in Fig.~\ref{fig:predictable}). An important contribution of this spatial evaluation of the \gls{oob} power is the ability to detect \gls{im} beams. These out-of-band beams can occur when multiple signals are fed to a nonlinear component, and in general are beamed towards more (different) directions than the in-band signals~\cite{arraybible}.

To evaluate the \gls{ib} signal quality, the \gls{evm} is calculated as follows:
\begin{equation}
    EVM \triangleq \frac{\left\| (\text{QAM}_\text{TX} - \text{QAM}_\text{RX})^2 \right\|}{\left\| \text{QAM}_\text{TX}^2 \right\|},
\end{equation}
with $\text{QAM}_\text{TX}$ the transmitted \gls{qam} symbols and $\text{QAM}_\text{RX}$ the received \gls{qam} symbols. $\left\|.\right\|$ is the  absolute value operator for complex signals. Another \gls{ib} quality evaluation can be performed by means of the constellation diagram of the received QAM symbols. This visualization clearly indicates whether or not the received symbols are interpretable after transmission. Similar to the \gls{oob} heatmap, an \gls{ib} heatmap is possible as well. These \gls{ib} metrics are applied for verification of the framework, it is not explicitly used in the analysis below.

\subsection{Grid of observers and users}
We divide the considered coverage in grid points, where each grid point is either an observer point or an intended user, allowing us to assess both \gls{ib} performance and \gls{oob} emissions at locations where users in adjacent may want to communicate. We thus sample the received power over a predefined area. The granularity of the grid is limited by computation resources. We have divided the area in 625 grid points, or 25~x~25, for example visible in Fig.~\ref{fig:predictable}.

\section{System parameters and scenario validation}\label{sec:analysis}
To analyze this multi variable and thus complex matter, we constraint some parameters and circumstances. We also describe in this section how we validated the framework with a scenario where the \gls{im} beams are predictable.

\subsection{General scenario parameters}
We here allocate some parameters and circumstances of the scenarios considered in the further analysis and presented results. The following parameters apply, unless specified otherwise: 
\begin{itemize}
    \item The \gls{bs} is equipped with 128 patch antennas in a \gls{ula} topology and is positioned at the center of the left edge of the considered area.
    \item The antenna spacing at the \gls{bs} is $0.5\lambda$ with $\lambda$ the wavelength of the center frequency of 3.5~GHz.
    \item The \gls{bs} is standard located at a height of 29~m. Exception to 50~m is made in view of creating \gls{los} conditions.
    \item The \glspl{ue} are equipped with a single isotropic antenna, located at a height of 1.5~m.
    \item The map area is the KU~Leuven Technology Campus in Ghent\footnote{\url{https://iiw.kuleuven.be/english/ghent/contact}}, size 440~m~x~444~m. The site includes relatively high buildings as well as some open space and is a good representation of an urban environment.
    \item The \gls{pa} \gls{bo} is 7~dB.
    \item Precoding is realized with \gls{mr}.
\end{itemize}

\subsection{LoS and nLoS scenarios}\label{sec:losnlos}
The scenarios in an urban environment inherently contain multipath components because of scattering, reflection, diffraction, etc. We expect these \textit{\gls{nlos} scenarios} to be favorable for distributing \gls{oob} emission spatially. To create a pure \textit{\gls{los} scenario} of which we expect it to follow theoretical predictions, we put both the \gls{bs} and the \glspl{ue} at an elevated height of 50~m. With that height, we are certain that no building will interfere or block the signal. It should be noted that an accidental \gls{los} beam between a \gls{ue} and the \gls{bs} could occur in a \gls{nlos} scenario. However, we look at the bigger picture of the complete area.

\subsection{Framework validation on predicted scenario}\label{sec:fact-check}
To validate the correctness of the results of the simulation framework, a scenario with theoretically predictable outcome is simulated. We assume a bandlimited system such that only $2f_1-f_2$ and $2f_2-f_1$ frequencies will be radiated in the neighbouring band. The simple case of two users in a \gls{los} is considered, where the angle towards the \gls{bs} is denoted as $\theta_i$. The boresight of the antenna array, perpendicular on the array, is the reference angle of $0^{\circ}$. The angles of the \gls{im} beams $\theta_{IM,A}$ and $\theta_{IM,B}$ can be calculated as:
\begin{equation}
    \theta_{I M, A}=\sin ^{-1}\left(\frac{2 f_{1} \cdot \sin \theta_{1}-f_{2} \cdot \sin \theta_{2}}{2 f_{1}-f_{2}}\right)
\end{equation}
\begin{equation}
    \theta_{I M, B}=\sin ^{-1}\left(\frac{2 f_{2} \cdot \sin \theta_{2}-f_{1} \cdot \sin \theta_{1}}{2 f_{2}-f_{1}}\right).
\end{equation}
Assuming for an approximate estimate of these angles that $f_1\approx f_2$, these formulas simplify to:
\begin{equation}
    \theta_{I M, A} \approx \sin ^{-1}\left(2 \sin \theta_{1}-\sin \theta_{2}\right)
\end{equation}
\begin{equation}
    \theta_{I M, B} \approx \sin ^{-1}\left(2 \sin \theta_{2}-\sin \theta_{1}\right)
\end{equation}


\begin{figure}[!htb]
    \centering
    \includegraphics[width=0.8\linewidth]{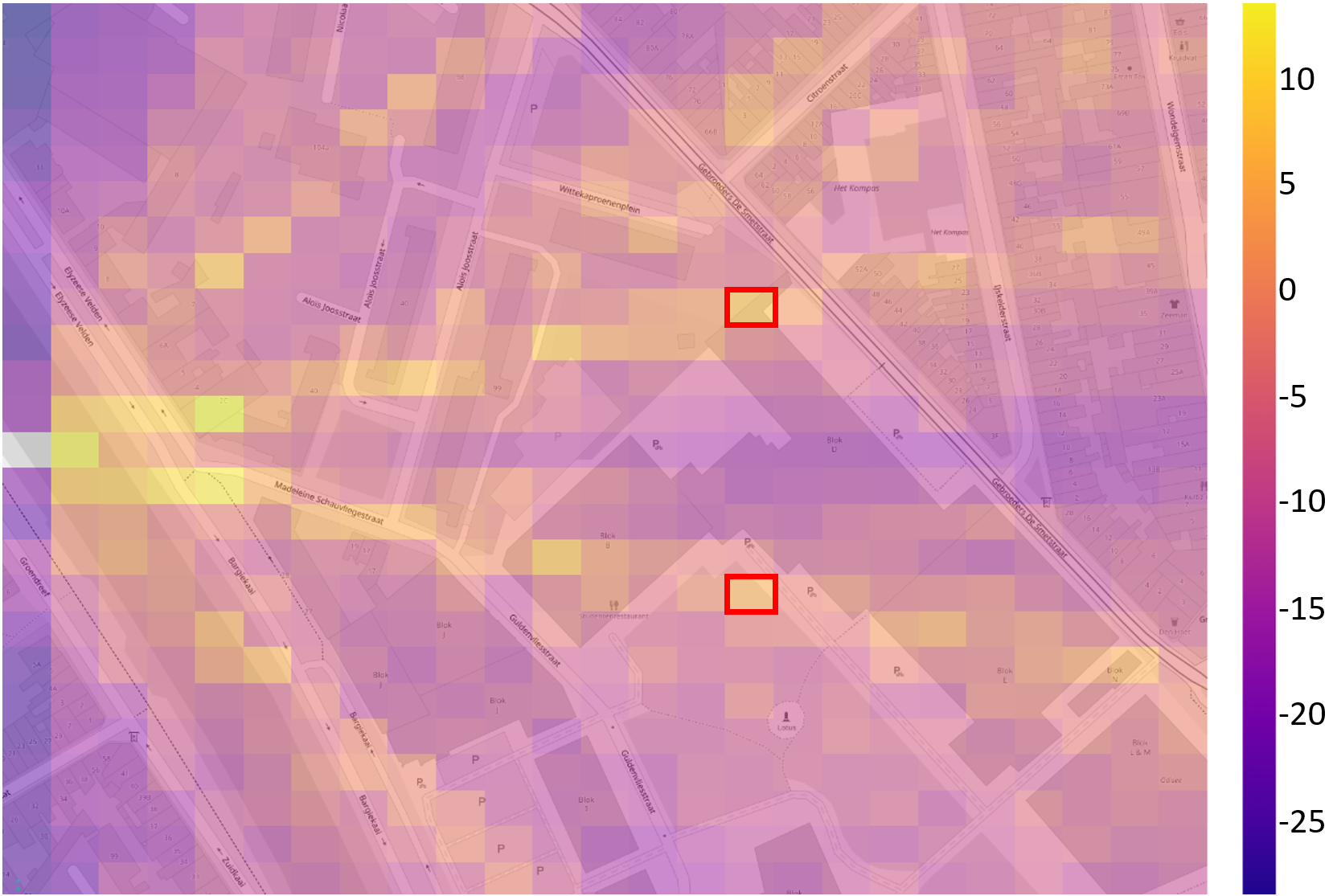}
    \caption{\gls{oob} heatmap of a \gls{los} experiment with two users on $15^{\circ}$ and $-15^{\circ}$, indicated by the red squares, with \gls{bs} at the center of the left edge.
    }
    \label{fig:predictable}
\end{figure}

The experiment is carried out with $\theta_1= -\theta_2 = 15^{\circ}$, which should yield in intermodulation beams in the directions $\theta_{IM,A} = -\theta_{IM,B} \approx 50,94^{\circ}$. Fig.~\ref{fig:predictable} is the \gls{oob} heatmap of this scenario with the \gls{bs} \gls{ula} antenna array centered on the left side of the area. We clearly distinguish two beams to the users at $\theta_1$ and $\theta_2$, and two \gls{im} beams at about $50^{\circ}$ and $-50^{\circ}$. This experiment validates the simulation framework with an outcome that was theoretically predicted.

\section{Results}\label{sec:results}
\begin{figure}[!htb]
    \centering
    \includegraphics[width=0.85\linewidth]{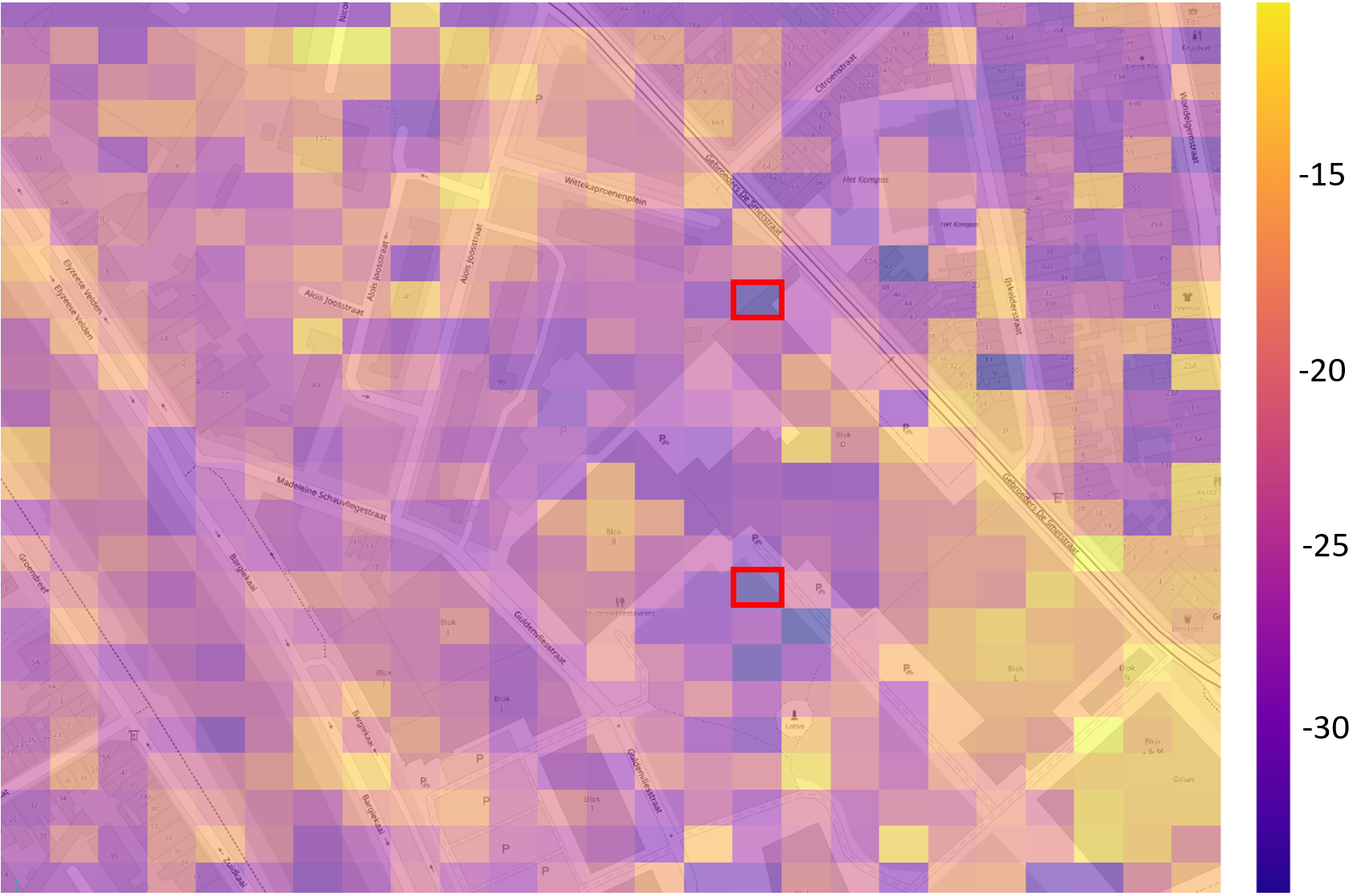}
    \caption{\gls{oob} heatmap of a \gls{nlos} experiment with two users with $\theta_1= 15^{\circ}$ and $\theta_2= -15^{\circ}$, indicated by the red squares, with \gls{bs} at the center of the left edge.}
    \label{fig:predictablenlos}
\end{figure}

In Section~\ref{sec:fact-check}, we investigated a specific predictable scenario under \gls{los} conditions, as described in Section~\ref{sec:losnlos}. We here extend the analysis for handpicked scenarios that we anticipate to be representative for diverse cases of special interest. The first case (not shown in view of space constraints) is a single user \gls{los} scenario, where the simulations confirm the theoretically predicted results that no intermodulation beams are generated, and the \gls{oob} emissions receive array gain similar to the \gls{ib} signals.
A next interesting case, is the \gls{nlos} version of the scenario in Section~\ref{sec:fact-check}, for which the resulting \gls{oob} heatmap is shown in Fig.~\ref{fig:predictablenlos}. A completely different spatial distribution pattern of the \gls{oob} is observed. 
It is important to notice that the color scale has shifted down, meaning that the intensity of \gls{oob} emissions has decreased, as also theoretically predicted for multi-path conditions. Still, the observed \gls{oob} shows high variations with up to 20~dB over different locations. 

\begin{figure}[!htb]
    \centering
    \includegraphics[width=0.85\linewidth]{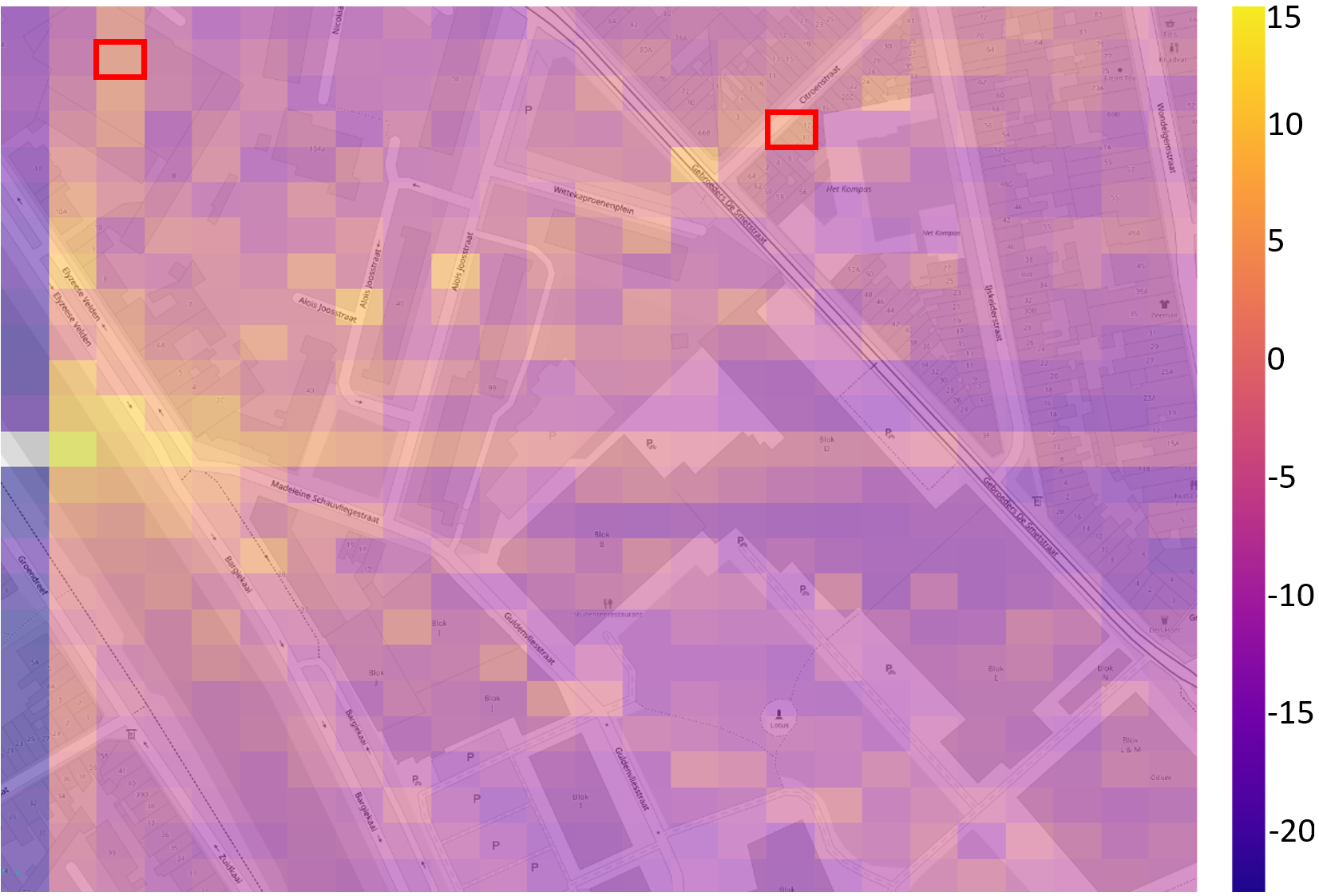}
    \caption{\gls{oob} heatmap of a \gls{los} experiment with two users with $\theta_1= 80^{\circ}$ and $\theta_2 =29.5^{\circ}$, indicated by the red squares, with \gls{bs} at the center of the left edge.}
    \label{fig:80-295-los}
\end{figure}

We conduct a similar \gls{los} and \gls{nlos} experiment, now with $\theta_1= 80^{\circ}$ and $\theta_2 =29.5^{\circ}$. Then, we expect an \gls{im} beam in the `favorite' direction of the antenna array, namely at $0^{\circ}$. The result confirms this as demonstrated in Fig~\ref{fig:80-295-los}. In case highly directive antenna elements are used, this could lead to significant \gls{oob} emissions in this direction with respect to power transmitted in the directions of the intended users.

\begin{figure}[!htb]
    \centering
    \includegraphics[width=0.85\linewidth]{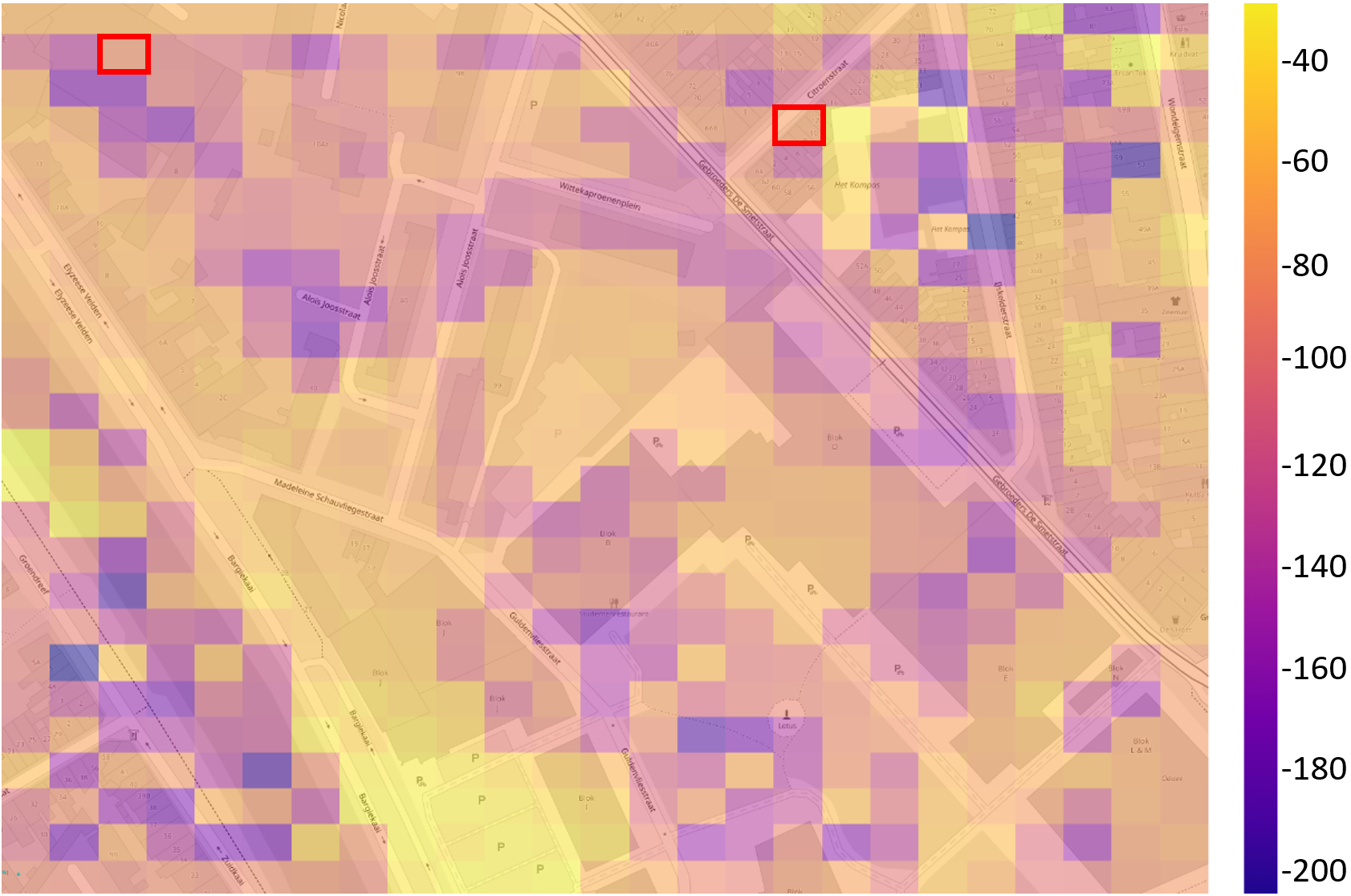}
    \caption{\gls{oob} heatmap of a \gls{nlos} experiment with two users with $\theta_1= 80^{\circ}$ and $\theta_2 =29.5^{\circ}$, indicated by the red squares, with \gls{bs} at the center of the left edge.}
    \label{fig:80-295-nlos}
\end{figure}

The corresponding experiment in \gls{nlos} conditions yields a much more scattered result, as seen in Fig.~\ref{fig:80-295-nlos}. These \gls{nlos} results confirm that more multi-path components lower coherent combining of \gls{oob} components from the many antennas in massive MIMO transmission. Still, the spatial distribution of the \gls{oob} emissions is far from uniformly distributed as theoretically predicted for i.i.d. Rayleigh fading channels. We have also assessed \gls{nlos} scenarios with more than $10$ users randomly distributed over the considered area, providing very similar results.

\begin{figure}
    \centering
    \includegraphics[width=\linewidth]{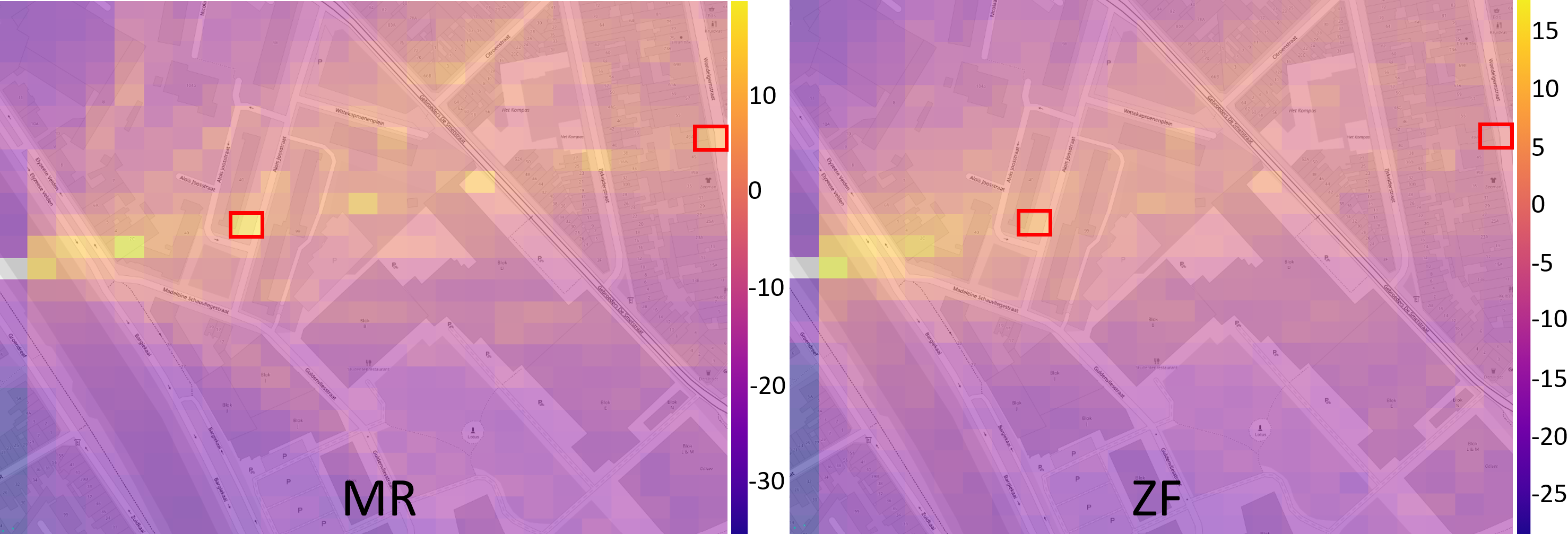}
    \caption{OOB heatmap for MR vs ZF precoding for a situation with two users on the same \gls{los}. } 
    \label{fig:MRvsZF}
\end{figure}

A last investigated scenario compares \gls{mr} and \gls{zf} precoding with two users on the same \gls{los} when seen from the \gls{bs}. It is known that \gls{mr} precoding will result, in this setting, in a bad \gls{evm} due to considerable multi-user interference distortion, while the \gls{zf} can still achieve good user separation. The results of the \gls{oob} analysis, depicted in Fig.~\ref{fig:MRvsZF}, demonstrate that the \gls{zf} is also better from this perspective, as it clearly spreads the distortion more uniformly in space. It should be noted that in practical deployments a regularised \gls{zf} precoding is preferred.

\section{Discussion of results in a regulatory context}\label{sec:regulations}
\begin{figure}
    \includegraphics[width=0.85\linewidth]{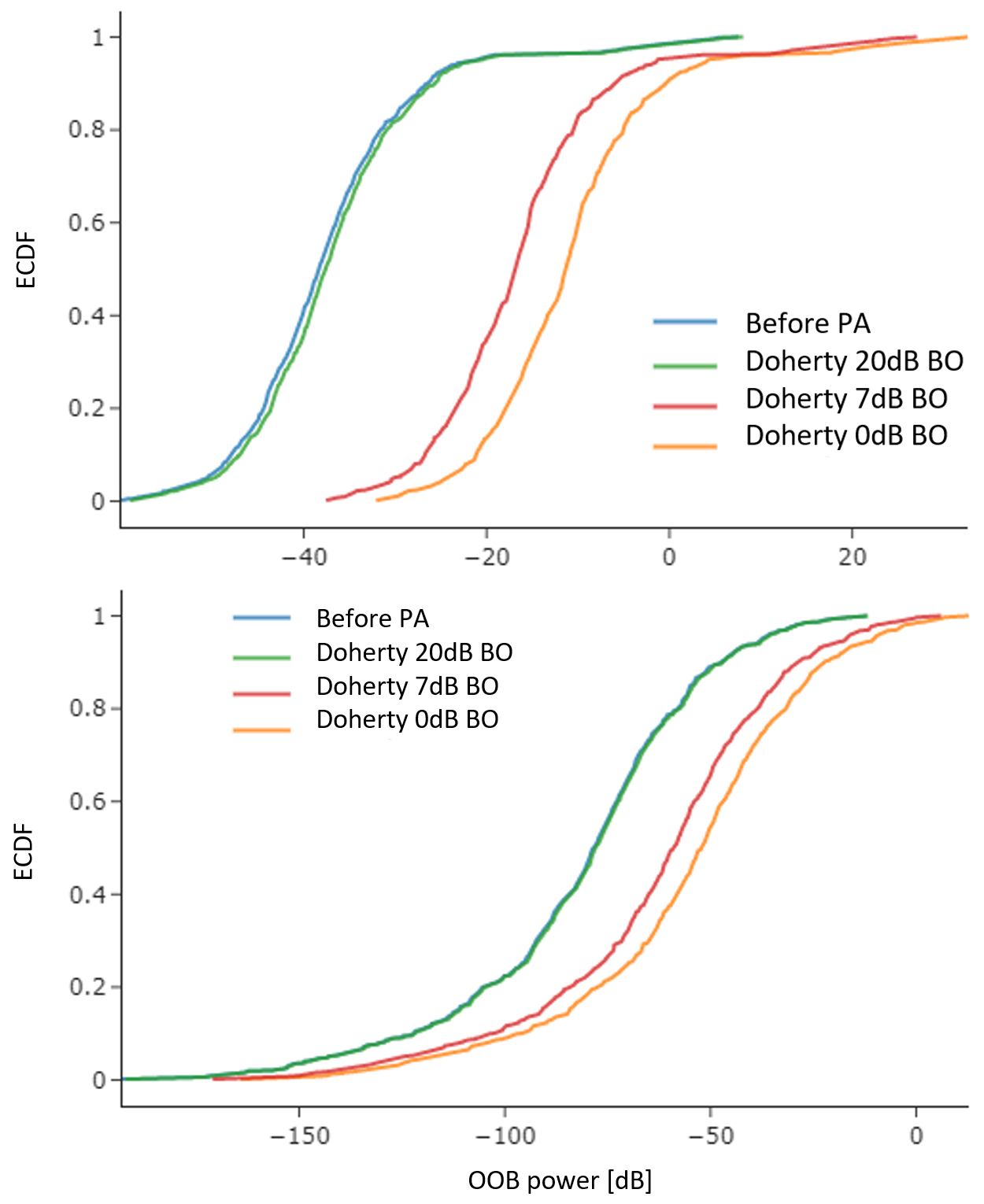}
    \caption{ECDF of \gls{los} single user (upper) and \gls{nlos} 4 user case (lower). Notice the different x axes because of the large spread of \gls{oob} power and the overlap of the green and blue curve in the figure at the bottom.}
    
    \label{fig:ecdf}
\end{figure}

In the context of \gls{oob} emissions, the legacy \gls{aclr} regulations are the so-called \textit{conducted measurements}. In that case, the \gls{aclr} measured at the \gls{arp} must be below a predefined threshold. The biggest question here is what this ARP should be in the case of massive MIMO. There is the consensus that we should not be looking at each antenna element separately but all of them combined, thus the antenna array rather than its elements. However, as \cite{Mollen2018} states, this type of measuring ignores the \gls{ag} and the effect it could have on the OOB radiation. A second type of measurements are the so-called \textit{over-the-air measurements}, where the measurements are done at selected positions around the \gls{bs}. Our \gls{oob} heatmaps offer a clear visualization of this second type of measurement. The results demonstrate that \gls{oob} is far from uniformly distributed in space. One may argue that conducting measurements \textit{everywhere} and for \textit{numerous user combinations} is required to guarantee that regulations are violated \textit{nowhere and never}. The theoretical analysis and simulation-based assessment can help to predict and provoke worst case \gls{oob} scenarios, as we have done in our study.\\
Moreover, one may also question the sustainability of current regulations. Indeed, setting the back-off of the \glspl{pa} at a level such that regulations will nowhere and never be violated may result in a low efficiency, while compliance is reached in most places also when working closer to saturation at higher efficiency. To this end, we depict \gls{ecdf} curves in Fig.~\ref{fig:ecdf} for the \gls{oob} power measured at all observer locations for the two distinct cases: a situation with a single user in \gls{los} on the one hand, and four users in \gls{nlos} on the other hand. First of all, it shows the large variations in space. Also, the considerable different \gls{ecdf} for different user scenarios suggest that in scenarios that may require a relative high output power to achieve good in-band performance (multiple users, \gls{nlos}), one may afford to operate the \glspl{pa} closer to saturation in a more efficient regime.

\section{Conclusions and future work}\label{sec:conclusion}
We investigated \gls{oob} distortion in a massive \gls{mimo} environment under realistic conditions by means of a simulation framework with integrated \gls{rts} channels. The performance evaluation metrics allow us to objectively analyze different scenarios. The framework was verified by fact-checking the simple scenario of two users in a \gls{los} environment with predictable \gls{im} beams.\\
We can conclude that the spatial distribution of \gls{oob} in realistic scenarios is not well predicted by theoretical analysis where specific simplified channel models are applied. Despite multi-path environments and multi-user scenarios, the results don't show uniformly distributed \gls{oob}. This reveals that the assumption of non-correlated distortions terms is also in these cases over-simplifying. The environment itself might not resolve the \gls{oob} violations. Significant beamforming gain in the \gls{oob} radiation is observed in \gls{los} as well as \gls{nlos} scenarios.
Our experience is that the \gls{rts} software is able to capture the complexity of the matter.\\
We comment that conventional regulations need revision for massive \gls{mimo} environments.
The evident next step is to find solutions to the problematic situations where \gls{oob} distortions exceed the regulations. These could involve dedicated user scheduling and adequate precoding approaches.

\section*{Acknowledgment}
We would like to thank NVIDIA for providing the GPU that greatly accelerated our simulations.

\bibliographystyle{IEEEtran}
\bibliography{references}

\begin{thebibliography}{10}
\providecommand{\url}[1]{#1}
\csname url@samestyle\endcsname
\providecommand{\newblock}{\relax}
\providecommand{\bibinfo}[2]{#2}
\providecommand{\BIBentrySTDinterwordspacing}{\spaceskip=0pt\relax}
\providecommand{\BIBentryALTinterwordstretchfactor}{4}
\providecommand{\BIBentryALTinterwordspacing}{\spaceskip=\fontdimen2\font plus
\BIBentryALTinterwordstretchfactor\fontdimen3\font minus
  \fontdimen4\font\relax}
\providecommand{\BIBforeignlanguage}[2]{{%
\expandafter\ifx\csname l@#1\endcsname\relax
\typeout{** WARNING: IEEEtran.bst: No hyphenation pattern has been}%
\typeout{** loaded for the language `#1'. Using the pattern for}%
\typeout{** the default language instead.}%
\else
\language=\csname l@#1\endcsname
\fi
#2}}
\providecommand{\BIBdecl}{\relax}
\BIBdecl

\bibitem{danve}
S.~R. {Danve}, M.~S. {Nagmode}, and S.~B. {Deosarkar}, ``{Energy Efficient
  Cellular Network Base Station: A Survey},'' in \emph{2019 IEEE Pune Section
  International Conference (PuneCon)}, 2019, pp. 1--4.

\bibitem{bjornson2014}
E.~{Bj\"ornson}, J.~{Hoydis}, M.~{Kountouris}, and M.~{Debbah}, ``{Massive MIMO
  Systems With Non-Ideal Hardware: Energy Efficiency, Estimation, and Capacity
  Limits},'' \emph{IEEE Transactions on Information Theory}, vol.~60, no.~11,
  pp. 7112--7139, 2014.

\bibitem{Desset}
C.~Desset and L.~Van~der Perre, ``{Validation of low-accuracy quantization in
  massive MIMO and constellation EVM analysis},'' in \emph{2015 European
  Conference on Networks and Communications}, 2015, pp. 21--25.

\bibitem{yadav}
S.~P. {Yadav} and S.~C. {Bera}, ``{Nonlinearity effect of high power amplifiers
  in communication systems},'' in \emph{2014 International Conference on
  Advances in Communication and Computing Technologies (ICACACT 2014)}, 2014,
  pp. 1--6.

\bibitem{ITU}
I.~T.~U. (ITU), ``{Unwanted emissions in the out-of-band domain},''
  \url{https://www.itu.int/dms_pubrec/itu-r/rec/sm/R-REC-SM.1541-6-201508-I!!PDF-E.pdf},
  2015, accessed: 2021-04-28.

\bibitem{Larsvdp2018}
E.~G. {Larsson} and L.~{Van der Perre}, ``{Out-of-Band Radiation From Antenna
  Arrays Clarified},'' \emph{IEEE Wireless Communications Letters}, vol.~7,
  no.~4, pp. 610--613, 2018.

\bibitem{Mollen2018}
C.~{Mollén}, E.~G. {Larsson}, U.~{Gustavsson}, T.~{Eriksson}, and R.~W.
  {Heath}, ``{Out-of-Band Radiation from Large Antenna Arrays},'' \emph{IEEE
  Communications Magazine}, vol.~56, no.~4, pp. 196--203, 2018.

\bibitem{arraybible}
R.~J. Mailloux, \emph{{Phased Array Antenna Handbook}}, 3rd~ed.\hskip 1em plus
  0.5em minus 0.4em\relax USA: Artech House, Inc., 2017.

\bibitem{Metis}
J.~Medbo, K.~B\"orner, K.~Haneda, V.~Hovinen, T.~Imai, J.~J\"arvelainen,
  T.~J\"ams\"a, A.~Karttunen, K.~Kusume, J.~Kyr\"ol\"ainen, P.~Ky\"osti,
  J.~Meinil\"a, V.~Nurmela, L.~Raschkowski, A.~Roivainen, and J.~Ylitalo,
  ``{Channel modelling for the fifth generation mobile communications},'' in
  \emph{The 8th European Conference on Antennas and Propagation (EuCAP 2014)},
  2014, pp. 219--223.

\bibitem{welch}
P.~Welch, ``{The use of fast Fourier transform for the estimation of power
  spectra: A method based on time averaging over short, modified
  periodograms},'' \emph{IEEE Transactions on Audio and Electroacoustics},
  vol.~15, no.~2, pp. 70--73, 1967.

\bibitem{mollen}
C.~{Mollén}, U.~{Gustavsson}, T.~{Eriksson}, and E.~G. {Larsson},
  ``{Out-of-band radiation measure for MIMO arrays with beamformed
  transmission},'' in \emph{2016 IEEE International Conference on
  Communications (ICC)}, 2016, pp. 1--6.

\end{thebibliography}

\end{document}